\documentclass[12pt]{article}
\usepackage{graphicx,amsmath}
\usepackage{units}

\newlength{\dinwidth}
\newlength{\dinmargin}
\renewcommand{\vec}[1]{\boldsymbol{#1}}

\def\lapproxeq{\lower .7ex\hbox{$\;\stackrel{\textstyle                                                    
<}{\sim}\;$}}                                                    
\def\gapproxeq{\lower .7ex\hbox{$\;\stackrel{\textstyle                                                    
>}{\sim}\;$}}                                                    
\def\be{\begin{equation}}                                                    
\def\ee{\end{equation}}                                                    
\def\bea{\begin{eqnarray}}                                                    
\def\eea{\end{eqnarray}}
\def\b{\vec{b}}
     
\def\q{\vec{q}}

\def\GeV{\rm GeV}

\def\sh{\hat s}
\def\sh2{{\hat s}^2}
\def\sd{$\sigma^{\rm D}_{{\rm low}M}~$}

\begin{document}
                                                    
\titlepage                                                    
\begin{flushright}                                                    
IPPP/18/42  \\                                                    
\today \\                                                    
\end{flushright} 
\vspace*{0.5cm}
\begin{center}                                                    
{\Large \bf Elastic and diffractive scattering at the LHC}\\

\vspace*{1cm}
                                                   
V.A. Khoze$^{a,b}$, A.D. Martin$^a$ and M.G. Ryskin$^{a,b}$ \\                                                    
                                                   
\vspace*{0.5cm}                                                    
$^a$ Institute for Particle Physics Phenomenology, University of Durham, Durham, DH1 3LE \\                                                   
$^b$ Petersburg Nuclear Physics Institute, NRC Kurchatov Institute, Gatchina, St.~Petersburg, 188300, Russia

\vspace*{1cm}                                                    
                                                    
\begin{abstract}                                                    

Inspired by the new TOTEM data on elastic $pp$ scattering at 13 TeV, we study the possibility to describe all the diffractive collider data ($\sigma_{\rm tot}, d\sigma_{\rm el}/dt$, $\rho\equiv$Re$A$/Im$A$, \sd) in a wide interval of energy (0.0625 to 13 TeV) in the framework of a two-channel eikonal model. We show that a satisfactory description can be achieved without an odd-signature (Odderon) exchange contribution. We consider the possible role of the QCD Odderon which may improve the description of $\rho$ and discuss the importance of the odd-signature term if the amplitude were to exceed the black disc limit. 
%The present TOTEM data exceed the limit by more than $3\sigma$ at impact parameter $b=0$.  
 
\end{abstract}                                                        
\vspace*{0.5cm}                                                    
                                                    
\end{center}

 \section{Motivation}
 TOTEM have recently published very detailed data on elastic proton-proton scattering at 13 TeV at the
 LHC,
covering the very low $t$ region\footnote{The behaviour of the cross section at very low $t$ samples Coulomb interference and allows a measure of the real part of the amplitude.} and up to the region of the diffractive dip and well beyond \cite{Antchev:2017yns,Antchev2}. The goal of this paper is to describe these data together with elastic and diffractive data at other collider energies. We use a two-channel eikonal model.  In addition to the dominant even-signature amplitude, we discuss the role of the odd-signature (Odderon) amplitude.  Moreover we study the present situation concerning information on low-mass diffractive dissociation. Recall that the multi-channel eikonal model is written, in the Good-Walker formalism~\cite{Good:1960ba}, in terms of diffractive eigenstates; and the experimental information on low-mass diffraction, $\sigma^D_{{\rm low}M} $, controls the relative contributions of the different diffractive eigenstates.
 
 Finally, we discuss the high-energy behaviour of the elastic amplitude for a central collision. That is, at impact parameter $b$~=~0.

\section{Description on the model  \label{sec:2}}
We use a two-channel eikonal model which is based on the two-particle unitarity equation,
\be
2{\rm Im} A_{\rm el}(b)~=~|A_{\rm el}(b)|^2 + G_{\rm inel}(b).
\label{eq:Ueq}
\ee
in impact parameter, $b$, space.
This accounts for the possibility of proton dissociation $p\to N^*$, not only in the final state, but also in an intermediate state. That is the model includes rescattering like $p\to N^* \to N^* \to p$ etc.  The beautiful and convenient way to accomplish this is to use the Good-Walker formalism \cite{Good:1960ba}, and to introduce diffractive eigenstates $| \phi_i \rangle$ which diagonalize the diffractive amplitude
\be
\langle\phi_i|A|\phi_k\rangle~=~A_{ik} ~=~A_i ~\delta_{ik}.
\ee
For each individual eigenstate the elastic amplitude is given by the one-channel eikonal expression
\be
\label{eik}
A(b)~=~i\left(1-e^{-\Omega(b)/2}\right).
\ee

The incoming `beam' proton wave function is written as a superposition of the diffractive eigenstates
\begin{equation}
|p\rangle~=~\sum a_i |\phi_i\rangle,
\label{eq:ai}
\end{equation}
and similarly for the incoming `target' proton. 
In this formalism the $pp$ elastic cross section is given by
\be
\frac{d\sigma_{\rm el}}{dt}~=~\frac{1}{4\pi}  \left| \int d^2b~e^{i\q_t \cdot \b} \sum_{i,k}|a_i|^2 |a_k|^2~(1-e^{-\Omega_{ik}(b)/2}) \right|^2,
\label{eq:dsel}
\ee
where $-t=q_t^2$, and the opacity $\Omega_{ik}(b)$ corresponds to  the interaction between states $\phi_i$ and $\phi_k$.
Also  the `total' low-mass diffractive cross section is of the form
 \be
\sigma_{\rm el+SD+DD}~=~  \int d^2b ~\sum_{i,k}|a_i|^2 |a_k|^2 ~\left|(1-e^{-\Omega_{ik}(b)/2}) \right|^2,
\ee
where SD includes the single dissociation of one or the other proton, and DD is the cross section for events where both protons dissociate.
So the low-mass diffractive dissociation cross section is
\be
\sigma^{\rm D}_{{\rm low}M}~=~\sigma_{\rm el+SD+DD}-\sigma_{\rm el},
\ee
where $\sigma_{\rm el+SD+DD}$ corresponds to all
possible low-mass dissociation caused by the dispersion of the Good-Walker
eigenstate scattering amplitudes. A more detailed description of the model is given in \cite{Khoze:2013dha}. 
 
As mentioned above, we use a two-channel eikonal, $i,k=1,2$.  Each eigenstate has its own coupling $v_i$ to the Pomeron, with its own $t$ dependence parametrised in the parametric form
\be
F_i(t)={\rm exp}(-(b_i(c_i-t))^{d_i}+(b_ic_i)^{d_i}),
\label{eq:ff}
\ee
where $c_i$ is added to avoid the singularity $t^{d_i}$ in the physical region of $t<4m^2_{\pi}$. Note that $F_i(0)=1$. The six parameters $b_i,~c_i,~d_i$, together with the intercept and slope of the pomeron trajectory are tuned to describe the elastic scattering data, paying particular attention to the observed energy behaviour of \sd, at all available collider energies, $\sqrt{s}$.

The opacity  $\Omega_{ik}$  corresponding to the scattering between eigenstates $\langle\phi_i|$ and $|\phi_k\rangle$ is given by one-Pomeron exchange
 \begin{equation}
 \Omega_{ik}(b)~=~\int\frac{d^2q_t}{4\pi^2}e^{i\vec q_t\cdot \vec
b}\Omega_{ik}(t=-q^2_t)\ ,
 \end{equation}
 with
 \begin{equation}
 \Omega_{ik}(t)~=~v_i F_i(t) v_k F_k(t)\left(\frac
s{s_0}\right)^{\alpha_{\rm P}(t)-1}
 \end{equation}
 and $s_0=1$ GeV$^2$.

\section{Low-mass proton dissociation \label{sec:3}}
 Note that if the amplitudes are identical, $A_i=A$, then the interaction will not destroy the coherence of the original proton wave function (\ref{eq:ai}).  Then the final state that we observe will be only the proton, while the probability of dissociation given by \sd will be zero. That is, a larger value of \sd indicates a larger dispersion between the amplitudes $A_i$.
 
 A model with a large number of Good-Walker components may account for different proton excitations and in this way describe $d\sigma/dM_X$,
 where $M_X$ is the mass of the system after the $p\to X$ dissociation. 
  In our $t$-channel eikonal analysis we use only one effective $N^*$ state, assuming that it includes all the excitations up to $M_X=3.4$ GeV, the mass value used by TOTEM collaboration~\cite{Antchev:2013haa} to separate proton dissociations into low- and high-mass states.\footnote{High-mass dissociation is described separately in terms of triple-Regge diagrams.}
 
 Experimentally the situation for measurements of \sd is far from clear. At the relatively low
\cite{Kaydalov:1971ta,Kaidalov:1979jz}
and ISR energies \cite{Baksay:1974mn,Webb:1974nb,Baksay:1976hb,deKerret:1976ze,Mantovani:1976vz}
 \be 
\sigma^D_{{\rm low}M}~\sim ~0.3~\sigma_{\rm el}~\sim ~2.5~{\rm mb},
 \ee 
while at 7 TeV TOTEM \cite{Antchev:2013haa} reported a much smaller value for the ratio \sd $/\sigma_{\rm el}$
\be 
\sigma^D_{{\rm low}M}~\sim ~2.6\pm 2.2~{\rm mb}~\sim~0.1~\sigma_{\rm el}.
\label{eq:9} 
\ee  
Recall that stronger absorptive corrections can decrease the ratio.

The situation at 13 TeV is not so evident. At the moment there are no TOTEM data for \sd. However we can compare the values of the inelastic cross sections measured by ATLAS \cite{Aaboud:2016mmw} and CMS \cite{Sirunyan:2018nqx} with the total and elastic cross section given by TOTEM \cite{Antchev:2017dia}.  A small complication is that ATLAS measure $\sigma_{\rm inel}=68.1\pm 0.6 \pm 1.3$ mb, using events where at least one particle carries a momentum fraction $\xi>10^{-6}$. This corresponds to $M_X > 13$ GeV. On the other hand, CMS use the CASTOR detector to cover the region down to $\xi>10^{-7}$ on one side of the interaction point. In other words CMS collect events for all processes except for the possible dissociative   $pp\to X+Y$, with $M_Y<4.1$ GeV on one side and $M_X<13$ GeV on the other side \cite{Sirunyan:2018nqx}. If we compare the two CMS results then we can estimate
\bea
\frac{d\sigma^D}{dln(M^2_X)}~&=&~\frac{\sigma_{\rm inel}(\xi_Y>10^{-7},\xi_X>10^{-6})-\sigma_{\rm inel}(\xi_X,\xi_Y>10^{-6})}{2ln(M_X/M_Y)} \\
&=&\frac{68.6-67.5}{2ln(13/4.1)}~\simeq ~0.48~{\rm mb}.
\label{eq:0.5mb}
\eea 
 This CMS number is in agreement with the ATLAS data \cite{Aad:2012pw} at 7 TeV on $d\sigma /d\Delta \eta^F$ for events with rapidity gaps and with the theoretical estimates of \cite{Ryskin:2011qe,Khoze:2013dha}.  Taking (\ref{eq:0.5mb}), we can evaluate the cross section of events with $M_{X,Y}<3.4$ GeV to be equal to 70.1 mb. Thus the cross section of dissociation up to the canonical $M_X=3.4$ GeV is
 \bea
 \sigma^D_{{\rm low}M}~&=&~\sigma_{\rm tot}-\sigma_{\rm el}-\sigma_{\rm inel}(M_X>3.4) \\
 &=&110.6-31-70.1~\simeq~9~{\rm mb}.
 \label{eq:B}
 \eea

A comparison of (\ref{eq:9}) and (\ref{eq:B}) shows that the value of \sd increases about three times in the relatively small energy interval from 7 to 13 TeV. This is very strange. Within this rather small ln$s$ interval we expect the variation of \sd to be of the order of 0.5 mb. Note, however, that the estimate (\ref{eq:B}) is obtained from the difference of two large numbers coming from different experiments\footnote{If we replace $\sigma_{\rm inel}(\xi>10^{-6})=67.5\pm1.6$(lumi) mb~\cite{Sirunyan:2018nqx} by the ATLAS value
 of $68.1\pm 1.3$(lumi) mb ~\cite{Aaboud:2016mmw} then we find the bit smaller value of about 8 mb in (\ref{eq:B}).} with their own normalization uncertainties like $\pm 3.4$ mb for $\sigma_{\rm tot}$(TOTEM) and $\pm 1.6$ mb for $\sigma_{\rm inel}$(CMS).\footnote{It is also possible that the value of $ d\sigma^D/dln(M^2_X)$ may be a bit larger for a lower $M_X$ value, in particular due to secondary Reggeon contributions, see Fig.9 of \cite{Ryskin:2011qe}. This would enlarge $\sigma_{\rm inel}$ and therefore decrease \sd a little; though, however, less than 0.5 mb.}
 The results of the model description that we shall present in Section \ref{sec:6} give \sd=~5.0 and 5.4 mb at 7 and 13 TeV respectively. In view of the uncertainties just discussed above, the model values are consistent with   all data.

 \section{Real part of the elastic $pp$ amplitude   \label{sec:4}}
 At high energy the elastic scattering amplitude is dominantly imaginary.  The ratio Re$A$/Im$A$ is about 0.1 and the real part plays a very small role in the low $t$ region (except of the Coulomb interference).  Nevertheless for a detailed description of the present very precise data we must account for this contribution.  Therefore in (\ref{eq:dsel}) we have to keep the full complex opacity $\Omega_{ik}(b)$ in the formula for the elastic amplitude
 \be
 A_{ik}(b)~=~\left(1~-~e^{-\Omega_{ik}(b)/2}\right).
 \label{eq:Uik}
 \ee
For Pomeron exchange we have the even-signature factor 
\be
\eta_{\rm even}~=~[1+\rm exp(-i\pi\alpha_P(t))]
\ee
for the `even' part of the opacity $\Omega_{ik}$(b), where $\alpha_P(t)$ is the Pomeron trajectory. If we keep only the even-signature contribution, the real part of the elastic amplitude satisfies the usual dispersion relation and its value can be calculated at $t=0$ from the known total cross section. Indeed, up to collider energies of $\sqrt{s}=8$ TeV, the experimental results for $\rho\equiv$ Re$A$/Im$A |_{t=0}$ ~are consistent with those deduced from the dispersion relation for the even-signature amplitude (see, for example, ~\cite{Cudell:2002xe}). 

However, at 13 TeV the TOTEM collaboration \cite{Antchev:2017yns} have reported a measurement of $\rho=0.09-0.10$ with an uncertainty $\pm 0.01$ which is significantly lower than $\rho=0.135$ expected by the conventional COMPETE analysis \cite{Patrignani:2016xqp}. If this difference were to be explained in the even-signature approach, it would indicate a slower growth of the total cross section with $\sqrt{s}$ than that given by the COMPETE parametrization, as stated in \cite{Antchev:2017dia}. On the other hand, the TOTEM \cite{Antchev:2017dia} measured value of the total cross section at 13 TeV 
\be
\sigma_{\rm tot}=110.6\pm 3.4~{\rm mb}
\ee
 is even a bit larger than that given by COMPETE \cite{Patrignani:2016xqp}.

 \section{Odderon exchange}
 Another way to obtain a smaller value of $\rho$ is to include the odd-signature (Odderon) contribution in the opacity  $\Omega_{ik}$(b).  The odd-signature
  factor with $\alpha_{\rm Odd}$ close to 1 
\be
\eta_{\rm odd}~=~[1-\rm exp(-i\pi\alpha_{Odd}(t))]
\label{eq:etaodd}
\ee
gives an almost real contribution to the elastic amplitude.
The Odderon
is expected in perturbative  QCD~\footnote{QCD is the $SU(N=3)$ gauge theory which contains the spin=1 particle (gluon) and (for $N>2$) the symmetric colour tensor, $d^{abc}$. Due to these facts in perturbative QCD there {\em exists} a colourless $C$-odd $t$-channel state (formed from three gluons) with intercept, $\alpha_{\rm Odd}$, close to 1.}, 
see in particular \cite{Kwiecinski:1980wb,Bartels:1980pe,Bartels:1999yt}.
However the naive estimates show that its contribution is rather small;  
say, $\Delta\rho_{\rm Odd}\sim 1\mbox{mb}/\sigma_{\rm tot}\lapproxeq 0.01$~\cite{Ryskin:1987ya} at the LHC
energies. 
The discovery of the long-awaited, but experimentally elusive, Odderon would be 
very welcome news for the theoretical community.
 Indeed, there have been several 
attempts to prove its existence experimentally
(see, for example,  
\cite{Braun:1998fs,Ewerz,Block} for comprehensive reviews and references).
 
 It is important to note that the Odderon contribution must be included in the opacity  $\Omega_{ik}$(b),
 and not directly in the elastic amplitude, since (\ref{eq:Uik}) is the general form of the solution of the two-particle unitarity equation where $\Omega_{ik}$ includes the {\it full} two-particle irreducible component of the interaction amplitude. Provided we include the odd-signature contribution to $\Omega_{ik
}$(b) via (\ref{eq:Uik}) we automatically account for the absorptive effect caused by elastic rescattering.\footnote{More details on the inclusion of the Odderon can be found in \cite{Finkelstein:1989mf,Khoze:2018bus}.}

   \begin{figure} [htb]
\begin{center}
%\vspace*{-5.0cm}
\includegraphics[trim=0 0cm 0 7.6cm,scale=0.6]{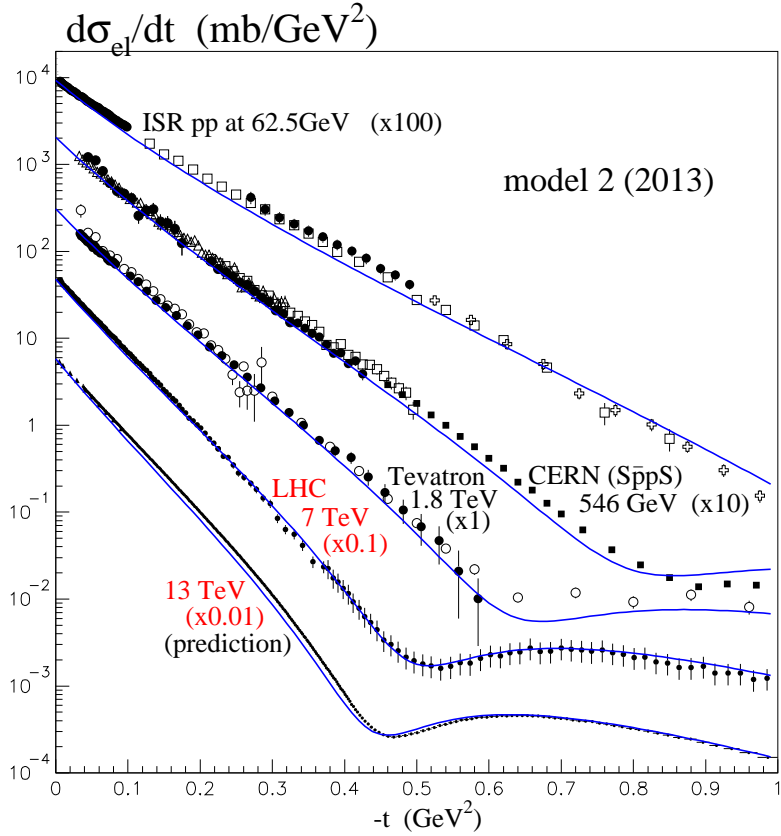}
\vspace{-0.0cm}
\caption{\sf The 2013 description of the $pp$ (and $p\bar{p}$) elastic data up to 7 GeV in model 2 of \cite{Khoze:2013dha}, together with the prediction for 13 TeV. The references for the data are given in \cite{Khoze:2013dha}; note that the Tevatron experiments cover data at 1.8 and 1.96 TeV. The recent TOTEM 13 TeV data \cite{Antchev:2017yns,Antchev2} are superimposed on the plot; they are hard to distinguish from the prediction, except for an interval about $t=-0.3 $ GeV$^2$ where they lie above. }
\label{fig:1}
\end{center}
\end{figure}

 \begin{figure} [htb]
\begin{center}
%\vspace*{-5.0cm}
%includegraphics[height=18cm]{ken-fig2.pdf}
\includegraphics[trim=0 0cm 0 7.6cm,scale=0.63]{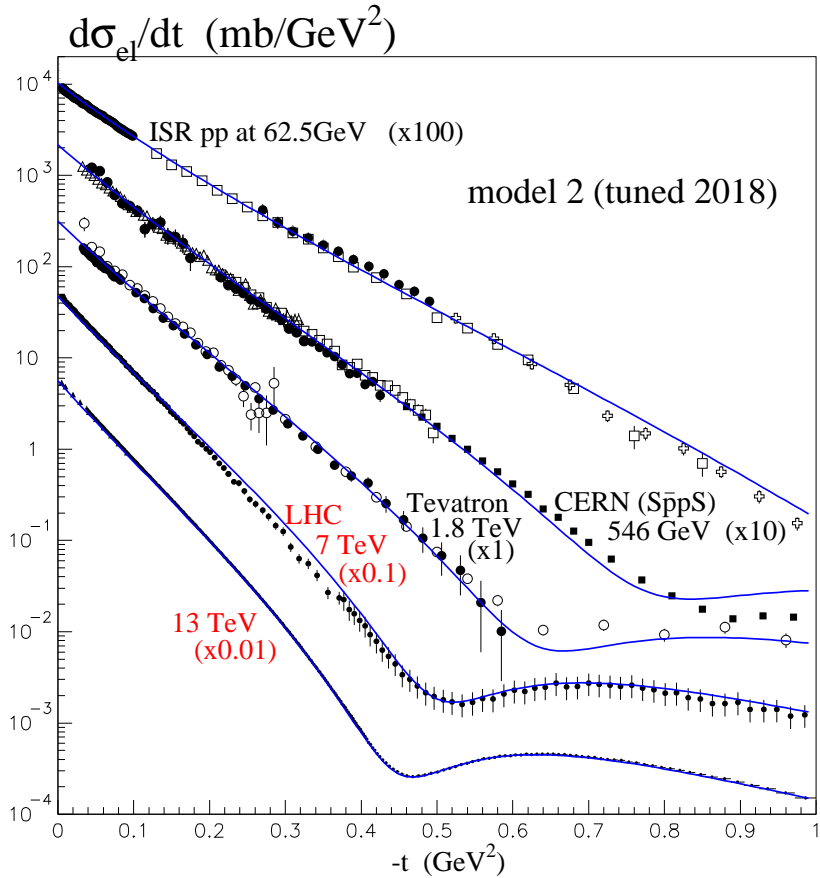}
\vspace{-0.0cm}
\caption{\sf As for Fig.~\ref{fig:1}, but now retuning the model to describe also the new TOTEM 13 TeV data \cite{Antchev2}. The description at 7 and 13 TeV is shown in more detail in the region of the diffractive dip by the continuous curves in Fig.~\ref{fig:3}.  }
\label{fig:2}
\end{center}
\end{figure}

\begin{figure} [htb]
\begin{center}
%\vspace*{-5.0cm}
\includegraphics[trim=0 0cm 0 7.6cm,scale=0.6]{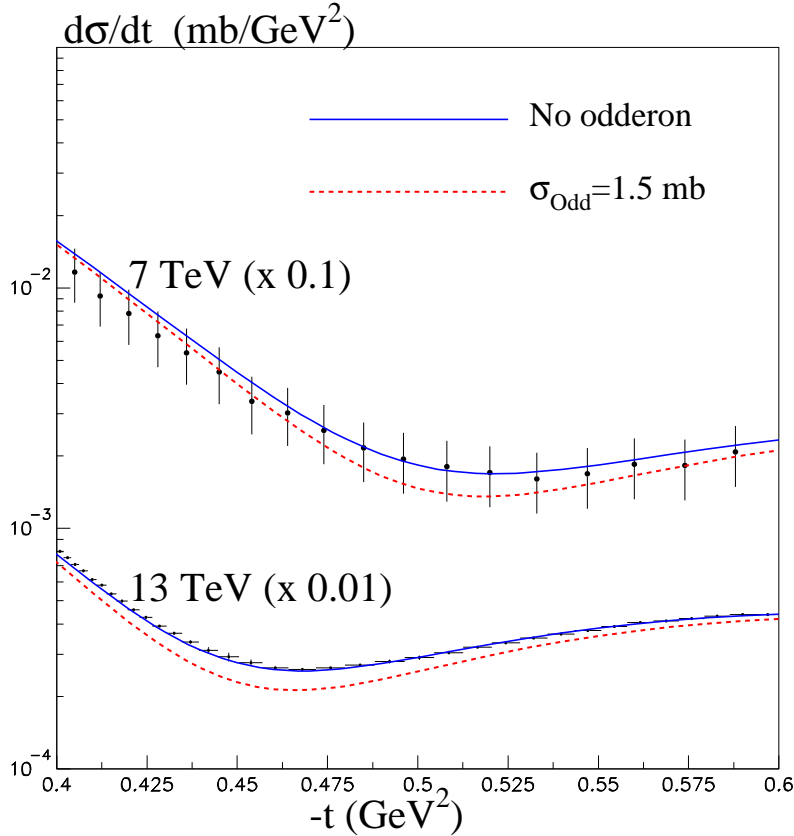}
\vspace{-0.0cm}
\caption{\sf The description of the 7 and 13 TeV data in the region of the diffractive dip enlarged from Fig.~\ref{fig:2}. In addition, the dashed curves show 
the effect of the Odderon if its contribution with $\sigma_{\rm Odd}=1.5$ mb
were to be included with the even-signature amplitude corresponding to the model used to obtain Fig.2.}
\label{fig:3}
\end{center}
\end{figure}
 
\begin{figure} [htb]
\begin{center}
%\vspace*{-5.0cm}
\includegraphics[trim=0 0cm 0 7.6cm,scale=0.61]{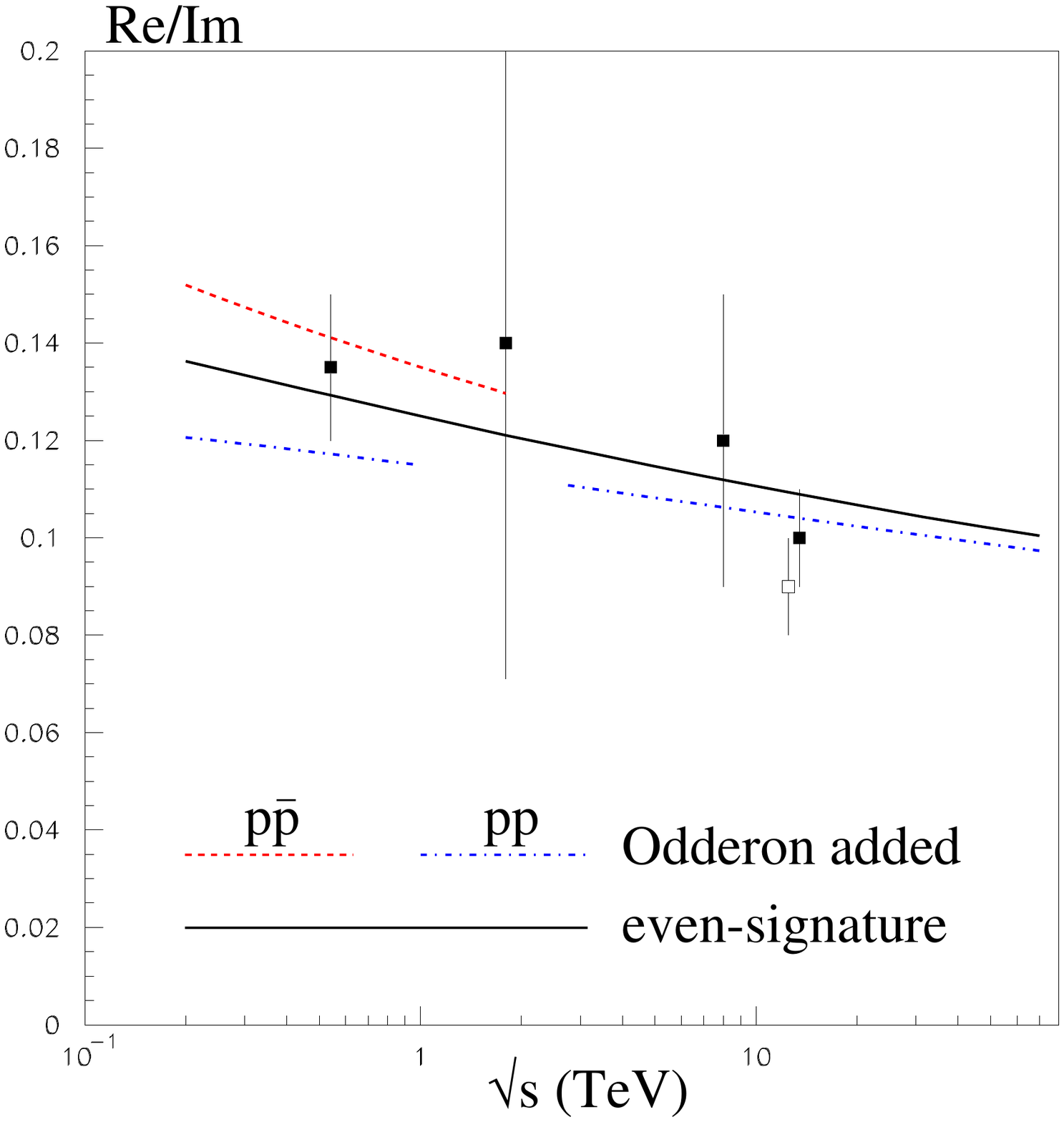}
\vspace{-0.0cm}
\caption{\sf The energy dependence of the $\rho=$Re$A$/Im$A$ ratio. The data are taken 
 from~\cite{Antchev:2017yns,Antchev:2016vpy,Amos:1991bp,Haguenauer:1993kan}; the first two data points correspond to $p\bar{p}$ scattering and the last two to $pp$ scattering. The values of $\rho$ given by the model are shown by the solid curve. The dashed ($p\bar{p}$) and dot-dashed ($pp$) curves correspond to an alternative behaviour of $\rho$ obtained from a `global' description of diffractive data which include a QCD Odderon contribution calculated as described in the text. }
\label{fig:4}
\end{center}
\end{figure}

\section{Results for the model description of the data \label{sec:6} } 
Using the two-channel eikonal model with a small set of parameters, we attempt to describe all the diffractive data ($\sigma_{\rm tot},~d\sigma_{\rm el}/dt, ~\rho\equiv $Re$A/$Im$A,$ \sd) over a wide range of collider energies (from $\sqrt{s}~=$ 0.0625 to 13 TeV) and a large interval of $t$ from 0 up to 1 GeV$^2$. The data correspond to more than four orders of magnitude variation for  
 $ d\sigma_{\rm el}/dt$.
 
 In the model the proton is described by a superposition of two diffractive eigenstates, ({\ref{eq:ai}), with form factors parametrised as in ({\ref{eq:ff}) and with coupling to Pomeron exchange given by
 \be
 v_{1,2}~=~\sqrt{\sigma_0}~(1\pm \gamma).
 \ee
 The Pomeron trajectory is parametrised as
 \be
\alpha_P(t)=1+\Delta+\alpha'_P t.
\ee
In addition to the constant slope, $ \alpha'_P$, of the Pomeron trajectory, we insert the $\pi$-loop contribution as proposed in \cite{Anselm:1972ir}, implemented as in \cite{Khoze:2000wk,Khoze:2014nia}. The parameter $\Delta$ embodies the BFKL effects which give $\Delta \sim 0.2-0.3$ and the renormalization
\footnote{Note that multi-Pomeron diagrams, in particular fan diagrams, also have an affect on the `effective' Pomeron form factor, (\ref{eq:ff}), tuned to describe the data. We do not explicitly include these multi-Pomeron diagrams in our simplified two-channel eikonal model in order to keep a clear physical structure of the interaction amplitude. The possible effects of these multi-Pomeron diagrams is allowed for by the renormalized parameters of the Pomeron trajectory and the Pomeron Good-Walker eigenstate couplings.} caused by the Pomeron loop insertions which decrease the resulting values of $\Delta$.
 
The original implementation of this model \cite{Khoze:2013dha}  described all the diffractive data existing up to 2013 in terms of only even-signature exchange. That is only the Pomeron contribution to $\Omega_{ik}(b)$. At that time there were a few local $\chi^2$ minima in parameter space corresponding to equally good descriptions of the data.  In fact we found, and presented \cite{Khoze:2013dha}, four versions of the model which gave good descriptions of all the elastic data available up to 7 TeV.  In Fig. \ref{fig:1} we reproduce the 2013 description of the elastic data up to 7 TeV given by version 2 of the model, which also showed the prediction made for $d\sigma_{\rm el}/dt$ at 13 TeV.  In addition we have superimposed on this figure the recent TOTEM measurement made at 13 TeV. The agreement at 13 TeV is surprisingly good. So what is the problem? The deficiency is in values of \sd. Version 2 of the model has values of \sd which are too small and no longer in agreement with the experimental information on \sd discussed in Section~\ref{sec:3}. The version 2 values of the cross section for low-mass diffractive dissociation are \sd = 1, 2.8, 3.1 mb at 0.0625, 7, 13 TeV respectively, which are too small.  However, if we  
retune the values of the parameters of version 2 (in particular, enlarging the value of $\gamma$ and make small adjustments to the form factors of the diffractive eigenstates) then we obtain the equally good description of the elastic data shown in Fig.~\ref{fig:2}, together with
\be
\sigma^D_{{\rm low}M}~=~2.35,~5.0,~ 5.4~~{\rm mb~~~~at}~~~~~0.0625,~7,~13~~{\rm TeV}
\ee
 respectively. Now the values do not contradict the experimental information discussed in Section~\ref{sec:3},  when we account for the experimental uncertainties.  The values of the parameters of the 2013 and 2018 descriptions of the data are shown in Table 1.  The observables as a function of energy corresponding to the present description of the data are shown in Table~\ref{tab:2}. It is informative to show in more detail in Fig.~\ref{fig:3}  (by continuous curves) the description of Fig.~\ref{fig:2} in the region of the diffractive dip for the 7 and 13 TeV data.
  \begin{table}[htb]
\begin{center}
\begin{tabular}{|l|c|c|c|c|}\hline
 &   2013 &  2018  \\ \hline
  $\Delta$ & 0.115  & 0.13 \\
  $\alpha'_P~(\GeV^{-2})$  & 0.11 & 0.052 \\
  $\sigma_0 $ (mb)  & 33 & 23\\
  $\gamma$ & 0.4  & 0.56\\
   \hline
  $|a_1|^2 $  & 0.25 &  0.505 \\
  $ b_1~(\GeV^{-2})$   & 8.0 & 10.0 \\
  $ c_1~(\GeV^2) $ & 0.18 & 0.233 \\
  $d_1$ & 0.63 & 0.462\\
  $ b_2~(\GeV^{-2})$   & 6.0 & 4.9  \\
  $ c_2~(\GeV^2) $ & 0.58 & 0.52 \\
  $d_2$ &  0.47  & 0.47\\
 \hline

\end{tabular}
\end{center}
\caption{\sf The values of the parameters in the two-channel eikonal fit to elastic $pp$ scattering data in which particular attention is paid to the value of \sd and to the behaviour of the GW eigenstates. The values of the parameters in the 2013 column correspond to version 2 of the original 2013 analysis \cite{Khoze:2013dha} (see Fig.~\ref{fig:1}), while the last column shows the values corresponding to the present description of the data (see Fig.~\ref{fig:2}). The first four rows give the values of parameters connected to the Pomeron trajectory and its couplings, and the last seven rows list the parameters which specify the Good-Walker diffractive eigenstates.}
\label{tab:1}
\end{table}

\begin{table}[htb]
\begin{center}
\begin{tabular}{|c|c|c|c|c|c|c|}\hline
 $\sqrt{s} $ & $\rho $   &    $\sigma_{\rm tot} $ & $\sigma_{\rm el} $ &  $B_{\rm el}(t=0)$  &   $\sigma^D_{lowM} $\\
 \hline
 
  (TeV)   &       &           mb    &     mb    &     (GeV$^{-2}$)  &  mb\\
\hline
    0.1  &    0.141   &    48.3   &     8.8   &    12.6   &     2.6\\
    0.546   & 0.129    &   64.6   &    13.8   &    14.8    &    3.5\\
   1.8  &    0.121   &    78.1   &    18.2   &    16.7    &    4.2\\
   7  &    0.113   &    95.5   &    24.1   &    19.1    &    5.0\\
   8   &   0.112   &    97.4   &    24.7    &   19.4    &    5.1\\
  13   &   0.109   &   104.2    &   27.1    &   20.4     &   5.4\\
 100  &    0.099   &   136.2   &    38.6   &    25.4     &   6.9\\
\hline

\end{tabular}
\end{center}
\caption{\sf The predictions of the elastic and diffractive observables resulting from the description of the presently available data.}
\label{tab:2}
\end{table}

 \subsection{The Odderon contribution  \label{sec:6.1}}
Before we discuss a possible Odderon contribution, we can see from Fig.~\ref{fig:4} that, even without an Odderon, the model produces a rather small value of  $\rho\equiv $Re$A/$Im$A=0.109$ at 13 TeV, more or less compatible with the 
recent TOTEM result \cite{Antchev:2017yns}. However what about our model value of $\sigma_{\rm tot}$ at 13 TeV? 
The present version of the model (constrained by the experimental information on low-mass proton dissociation, \sd, of Section~\ref{sec:3}) has a flatter energy behaviour of the total cross section. We slightly overestimate $\sigma_{\rm tot}$ at 62.5 GeV and underestimate $\sigma_{\rm tot}$ at 13 TeV, but are still in agreement with the data to within $1.5\sigma$.

What happens if we include an Odderon contribution?  In order not to introduce too many extra parameters, we use the same couplings for the odd-signature terms to the two different diffractive Good-Walker eigenstates. We parametrize the odd-signature amplitude as
\be
A_{\rm odd}(s,t)~=~s\sigma_{\rm Odd}~{\rm exp}({B_{\rm Odd}t})
\ee
where\footnote{Note that the C-odd and isospin=0 state does not couple to the pion. Thus the Odderon only feels the centre of the proton, and not the pion cloud. Therefore it is reasonable to assume that the Odderon slope, $B_{\rm Odd}$, is lower than that for the even-signature (Pomeron) amplitude.}
 $B_{\rm Odd}=6$ GeV$^{-2}$, and where the normalization corresponds to 
$\mbox{Im}A(t=0)~=~s\sigma_{\rm tot}$.
In other words we consider a QCD Odderon with intercept $\alpha_{\rm Odd}(0)~=~1$, as was obtained in \cite{Bartels:1999yt}, and normalization given by the parameter $\sigma_{\rm Odd}$.
The simplified lowest $\alpha_s$ calculation leads to $\sigma_{\rm Odd}~=~0.8$ mb \cite{Ryskin:1987ya}.
With such a small coupling of ${\cal O}$(1 mb) the Odderon is almost invisible in Fig.~\ref{fig:2}. 
 
Recall that, for $\alpha_{\rm Odd}(0)=1$, Odderon-exchange is real, see (\ref{eq:etaodd}). Thus we have essentially no interference term. The Odderon contribution only becomes visible in the dip region (see Fig~\ref{fig:3}) where the imaginary part of the even-signature amplitude vanishes; and in the region of very small $t$ where it interferes with the Coulomb ($\gamma$-exchange) term. 

Note that the Odderon decreases the value of $\rho\equiv $Re$A/$Im$A$ in $pp$ collisions, while simultaneously enlarging $\rho$ in $p\bar{p}$ collisions, see Fig.~\ref{fig:4}. Since the Odderon contribution must be added to the opacity and is screened in the full amplitude by Pomeron exchange it affects the value of $\rho$ at 13 TeV less than at 541 GeV where it was measured by the UA4 collaboration \cite{Haguenauer:1993kan} for $p\bar{p}$ scattering, again see Fig. \ref{fig:4}. In particular, setting the parameter $\sigma_{\rm Odd}~=~1.5$ mb we have $\Delta\rho~=~-0.005$ at 13 TeV and $\Delta\rho~=~+0.012$ at 541 GeV. 

Recall that we showed the role of the Odderon in the dip region in more detail in Fig.~\ref{fig:3}. We may conclude that even without the Odderon the model could be tuned to be consistent with the elastic data. However, a small Odderon comparable with the expectations of QCD may improve the agreement with the measurements on $\rho$, and not spoil the description of $d\sigma_{\rm el}/dt$ in the dip region, bearing in mind the uncertainties.

\section{Does the $pp$-amplitude exceed the black disc limit?  \label{sec:7}}
Naive predictions based on a Donnachie-Landshoff parametrization \cite{Donnachie:1992ny} show that the black disc limit is exceeded for central $(b=0)$ elastic $pp$ collisions at LHC energies.  That is, Im$A(s,b=0)>1$. It is therefore relevant to ask if the LHC data respect this limit.

Recall that the imaginary part of the high-energy elastic amplitude in impact parameter space is given by
\be
{\rm Im}A(b)~=~\frac{1}{(2\pi)^2}\frac{1}{2s}\int {\rm Im}A(t)~{\rm exp}(-ib\cdot q_t)~d^2 q_t
\ee
where $q_t=\sqrt{-t}$, and the values of Im$A(t)$ can be calculated directly from the data for the differential elastic cross section\footnote{The notation $\sigma^N$ means that to be precise we have to subtract from the measured elastic cross section the contributions caused by the pure Coulomb interaction and by Coulomb-nuclear interference.}
\be
|{\rm Im}A(t)|^2~=~\frac{16\pi s^2}{1+\rho^2}~\frac{d\sigma^N_{\rm el}}{|dt|},
\ee
with a small contribution ($\sim 1\%$) coming from $\rho^2$.  In this way we obtain
 \be 
 \label{A-b}
 \mbox{Im}A(b)=\int \kappa ~\sqrt{\frac{d\sigma^N_{\rm el}}{d|t|}~\frac{16\pi}{1+\rho^2}}~ J_0(bq_t)~\frac{q_tdq_t}{4\pi}
 \ee
 where $J_0$ is the Bessel function. Noting that Im$A(t)$ changes sign at the diffractive dip, we have $\kappa=+1$ or $-1$ for $|t|$ values below or above this point.

 There is some uncertainty since  we do not know the $t$ behaviour of the $\mbox{Re/Im}$ ratio $\rho$. On the other hand, the value of $\rho\sim 0.1$ is rather small, and assuming a flat behaviour ($\rho=const$ within the $t$ interval relevant for the integral (\ref{A-b})) we are able to calculate $A(b)$ with sufficient accuracy. 
 To obtain a rough estimate of Im$A(b=0)$ we may further simplify (\ref{A-b}) by assuming, in the relevant $|t|$ region, that the differential cross section is well described by a simple exponent $d\sigma^N_{\rm el}/dt\propto {\rm exp}(B_{\rm el}t)$. In such a case we get
 \be
 \label{A-exp}
 \mbox{Im}A(b=0)~=~\frac{\sigma_{\rm tot}}{4\pi B_{\rm el}}\ ,
 \ee
 which we evaluate using the published experimental data for $\sigma_{\rm tot}$ and $B_{\rm el}$. The results are presented in Table~\ref{tab:A}, where the errors have been added in quadrature.
 
  \begin{table}[htb]
\begin{center}
\begin{tabular}{|c|c||c| c|c|c|c|}\hline
$\sqrt s ~ {\rm TeV}$ & UA4:~~0.541 & LHC&~~ 2.76 &  7 & 8 & 13  \\ \hline
  Im$A(0)~$ & 0.84$\pm$ 0.025 & TOTEM& 1.01$\pm$ 0.043 & 1.01 $\pm$ 0.03 & 1.045$\pm$0.032&1.11$\pm$0.032 \\
  Im$A(0)~$ &  & ATLAS & & 0.988$\pm$0.02  & 0.996$\pm$0.015& \\
 \hline
 \end{tabular}
\end{center}
\caption{\sf The values of central amplitude, $A(b=0)$, obtained from (\ref{A-exp}) using the
 total cross sections and elastic slopes measured by UA4~\cite{Haguenauer:1993kan}, 
TOTEM~\cite{Antchev:2013haa,Antchev:2017dia,Antchev:2016vpy,Antchev:2013paa,Nemes:2017gut} 
and ATLAS-ALFA~\cite{Aad:2014dca,Aaboud:2016ijx} collaborations. Note that with present normalization the black disk limit corresponds to Im$A(b=0)=1$.}
\label{tab:A}
\end{table}
It is known that the proton-proton opacity, $\Omega$, increases with energy and correspondingly increases the value of $A(0)$.  Moreover it was claimed in ~\cite{Alkin:2014rfa} that already at 
$\sqrt s=7$ TeV the value of Im$A(b=0)>1$ exceeds the black disk limit $A=i$.
The surprising new result is that the value obtained from (\ref{A-exp}) for the TOTEM data at 13 TeV exceeds the limit by more than 3 standard deviations.  If confirmed, what would this mean?

 Recall that the expression (\ref{eik}), $A(b)=i(1-e^{-\Omega(b)/2})$, is the most general solution of the  unitarity equation (\ref{eq:Ueq}). That is, in order to obtain Im$A(b)>1$ we need to have $|{\rm Im}\Omega(b)|>\pi$. The opacity $\Omega=\Omega_{\rm even}+\Omega_{\rm odd}$ contains both the even and odd signature terms.      The imaginary contribution to $\Omega$ coming from the even-signature part is strongly limited by dispersion relations. It cannot exceed about 0.3$-$0.5. Such a large value, $|{\rm Im}\Omega(b)|>\pi$, can only come from an odd-signature contribution. For an exponential parametrization $A(t)\propto\exp(Bt)$ the value of $\Omega_{\rm odd}$ reads
 \be
 \label{omega}
 \Omega_{\rm odd}(b)=-i\eta_{\rm odd}\frac{\sigma_{\rm Odd}}{4\pi B}e^{-b^2/4B}\ .
 \ee
 In order to get $|{\rm Im}\Omega_{\rm odd}|>\pi$ with a reasonable slope $B=6$ GeV$^{-2}$ we would need the parameter  $\sigma_{\rm Odd}>90$ mb! This looks very unlikely\footnote{Formally such a large value of $\sigma_{\rm Odd}$ in a limited energy interval which includes 13 TeV does not violate unitarity. However, asymptotically as $s\to \infty$ the ratio Re$A$/Im$A$ must tend to 0, as was shown in \cite{Khoze:2018bus}.}. At present there is no model which can produce such a large real amplitude for high-energy $pp$-scattering.\footnote 
{Recall that within our approach we are unable to reproduce such a large Im$A(0)>1$ and the model would prefer a smaller value of $\sigma_{\rm tot}$ of about 105 mb at 13 TeV.}
If the value of Im$A(b=0)>1$ were to be confirmed, it would be an important hint in favour of a completely {\bf new} {\em strong} interaction beyond the Standard Model, which has never been observed before (for $\sqrt s\lapproxeq 1$ TeV) and reveals itself only in the LHC energy region.\footnote{Note however, that  the ATLAS-ALFA data at $\sqrt s=7$ and 8 GeV ~\cite{Aad:2014dca,Aaboud:2016ijx} are a bit below the black disk limit and all the previous results are consistent with Im$A(0)\leq 1$ within the error bars.}

But first we must question the simplified formula (\ref{A-exp}).  This approximation was acceptable at CERN-ISR energies where the position of the diffractive dip was at larger $|t|$, ($|t_{\rm dip}|\simeq 1.3$ GeV$^2$), and where the maximum value of $d\sigma_{\rm el}/dt$ after the dip never exceeds 
$10^{-6}d\sigma^N_{\rm el}(t=0)/dt$. However, at the LHC the dip occurs at much smaller $|t|\sim 0.5$ GeV$^2$ and the contribution of the negative amplitude, Im$A(t)<0$, to  the integral (\ref{A-b}) after the dip  is not negligible. Thus
 we must calculate the value of Im$A(b=0)$ more precisely based on (\ref{A-b}) and account for the fact that after the diffractive dip (i.e. for $|t|>0.47$ GeV$^2$ at 13 TeV) the imaginary part of the elastic amplitude changes sign. It turns out that at LHC energies the contribution after the diffractive dip 
  noticeably decreases the value obtained for Im$A(b=0)$.  The improved calculation gives
\be
\mbox{Im}A(b=0)~=~1.026\ .
\ee
 Bearing in mind a normalization uncertainty of about 3\% for $\sigma_{\rm tot}$,  this value is consistent with the statement that the amplitude {\em does not exceed black disk limit}.
 In fact we performed the calculation twice. First, we assumed a constant value $\rho=0.1$ independent of $t$,
 and second, we used the values of $\rho(t)$ given by the model described  in Sections \ref{sec:2}$-$\ref{sec:4}. The difference in Im$A(b=0)$ is negligible (less than 0.002).

 \section{Conclusions}
 We have considered the new TOTEM data \cite{Antchev2} on elastic $pp$ scattering at 13 TeV. We showed in Fig.~\ref{fig:2} that a satisfactory description of the $t$ distribution (and, in Fig~\ref{fig:4}, the TOTEM measurement of $\rho\equiv$Re$A$/Im$A$ \cite{Antchev:2017yns}) can be obtained in the framework of a two-channel eikonal model, even without the inclusion of an odd-signature (Odderon) contribution. However, the small addition of a QCD Odderon contribution may slightly improve the agreement with the data, 
  especially for the $\rho$ ratio. We emphasized, in Section~\ref{sec:7}, that {\em if} the value of the imaginary part of the amplitude at some impact parameter $b$ calculated from the 13 TeV data were to exceed the black disc limit
 this would be a strong argument in favour of a large odd-signature contribution. It is impossible to get Im$A(b)>1$ without a large odd signature
 term  
  (much larger than that expected from the perturbative QCD Odderon).

On other hand when we improved the calculation of Im$A(b)$, for the precise TOTEM 13 TeV data accounting in (\ref{A-b}) for the contribution from the large $|t|$ region (after the diffractive dip) where the imaginary part of the amplitude changes sign, we find 
\be 
{\rm Im}A(b=0)~=~1.026.
\ee
 Within the normalization error of about 3\% this is consistent with the `black disk limit' 
 Im$A(b)\le 1$.
   
 We emphasize that actually the main analysis of this paper was the description of all the diffractive data obtained for $pp$ (and $p\bar{p}$) collisions ($\sigma_{\rm tot},~d\sigma_{\rm el}/dt, ~\rho,$ \sd) over a wide range of collider energies (from $\sqrt{s}~=$ 0.0625 to 13 TeV) and a large interval of $-t$ from 0 up to 1 GeV$^2$, in terms of a two-channel eikonal model. In this `global' analysis, an overall satisfactory description of the data could be achieved either without, or with the inclusion of a small contribution from, a QCD Odderon. 
 
Note that the two-channel `global' description depends crucially on the experimental information on low-mass proton dissociation, \sd. The discussion in Section~\ref{sec:3} has required us to increase the values of \sd as compared to the values fitted in our earlier analyses. The consequence is that we have a flatter energy dependence of the total cross section -- we slightly overestimate the measured value of $\sigma_{\rm tot}$ at 62.5 GeV and underestimate the value measured at 13 TeV.
 That is, the {\it overall} description prefers a lower value of $\sigma_{\rm tot} \sim 105$ mb at 13 TeV, instead of the measured value $\sigma_{\rm tot} \sim 110.6\pm 3.4$ mb quoted by TOTEM 
 \cite{Antchev:2017yns}.  We await more precise experimental knowledge of \sd and 
 %even more detailed 
  further measurements of $\rho$ and $\sigma_{\rm tot}$.

 \section*{Acknowledgements}

MGR thank the IPPP at the University of Durham for hospitality. VAK acknowledges  support from a Royal Society of Edinburgh  Auber award.

\bibliographystyle{JHEP.bst}
\bibliography{Fvak}

\providecommand{\href}[2]{#2}\begingroup\raggedright\begin{thebibliography}{10}

\bibitem{Antchev:2017yns}
{\scshape TOTEM} collaboration, G.~Antchev et~al., \emph{{First determination
  of the $\rho$ parameter at $\sqrt{s}$ = 13 TeV – probing the existence of a
  colourless three-gluon bound state, CERN-EP-2017-335}}, {\emph{Submitted to:
  Phys. Rev.} (2017) }.

\bibitem{Antchev2}
F.~Nemes and F.~Ravera, \emph{contributions on behalf of the {TOTEM}
  collaboration, respectively, in the proceedings of the 4th {Elba} workshop on
  forward physics at {LHC} energy, 24-26 {May} 2018, to be published in a
  special issue of {Instruments}; and at the 134th {LHCC} meeting - open
  session, 30 {May} 2018}, .

\bibitem{Good:1960ba}
M.~L. Good and W.~D. Walker, \emph{{Diffraction disssociation of beam
  particles}}, \href{https://doi.org/10.1103/PhysRev.120.1857}{\emph{Phys.
  Rev.} {\bfseries 120} (1960) 1857}.

\bibitem{Khoze:2013dha}
V.~A. Khoze, A.~D. Martin and M.~G. Ryskin, \emph{{Diffraction at the LHC}},
  \href{https://doi.org/10.1140/epjc/s10052-013-2503-x}{\emph{Eur. Phys. J.}
  {\bfseries C73} (2013) 2503}
  [\href{https://arxiv.org/abs/1306.2149}{{\ttfamily 1306.2149}}].

\bibitem{Antchev:2013haa}
{\scshape TOTEM} collaboration, G.~Antchev et~al., \emph{{Measurement of
  proton-proton inelastic scattering cross-section at $\sqrt{s}$ = 7 TeV}},
  \href{https://doi.org/10.1209/0295-5075/101/21003}{\emph{EPL} {\bfseries 101}
  (2013) 21003}.

\bibitem{Kaydalov:1971ta}
A.~B. Kaydalov, \emph{{High-energy particle production and regge cuts in
  elastic n n, pi n, and k n scattering}}, {\emph{Sov. J. Nucl. Phys.}
  {\bfseries 13} (1971) 226}.

\bibitem{Kaidalov:1979jz}
A.~B. Kaidalov, \emph{{Diffractive Production Mechanisms}},
  \href{https://doi.org/10.1016/0370-1573(79)90043-7}{\emph{Phys. Rept.}
  {\bfseries 50} (1979) 157}.

\bibitem{Baksay:1974mn}
L.~Baksay et~al., \emph{{Observation of Diffraction Excitation in $p p \to$ ($p
  \pi^{+} \pi^{-)}$ x at the CERN - ISR and Tests of Limiting Fragmentation}},
  \href{https://doi.org/10.1016/0370-2693(75)90223-3}{\emph{Phys. Lett.}
  {\bfseries 53B} (1975) 484}.

\bibitem{Webb:1974nb}
R.~Webb et~al., \emph{{Production of Nucleon Resonances by Single Diffraction
  Dissociation at the CERN ISR}},
  \href{https://doi.org/10.1016/0370-2693(75)90614-0}{\emph{Phys. Lett.}
  {\bfseries 55B} (1975) 331}.

\bibitem{Baksay:1976hb}
L.~Baksay et~al., \emph{{Diffraction Dissociation in the Reaction $p p \to
  \Lambda K^+ p$ at the CERN ISR}},
  \href{https://doi.org/10.1016/0370-2693(76)90600-6}{\emph{Phys. Lett.}
  {\bfseries 61B} (1976) 405}.

\bibitem{deKerret:1976ze}
H.~de~Kerret et~al., \emph{{Experimental Results on Diffractive One Pion
  Production at the CERN ISR}},
  \href{https://doi.org/10.1016/0370-2693(76)90401-9}{\emph{Phys. Lett.}
  {\bfseries 63B} (1976) 477}.

\bibitem{Mantovani:1976vz}
G.~C. Mantovani, M.~Cavalli-Sforza, C.~Conta, M.~Fraternali, G.~Goggi,
  F.~Pastore et~al., \emph{{First Results on Diffraction Dissociation of
  Neutrons at the ISR}},
  \href{https://doi.org/10.1016/0370-2693(76)90125-8}{\emph{Phys. Lett.}
  {\bfseries 64B} (1976) 471}.

\bibitem{Aaboud:2016mmw}
{\scshape ATLAS} collaboration, M.~Aaboud et~al., \emph{{Measurement of the
  Inelastic Proton-Proton Cross Section at $\sqrt{s} = 13$  TeV with the
  ATLAS Detector at the LHC}},
  \href{https://doi.org/10.1103/PhysRevLett.117.182002}{\emph{Phys. Rev. Lett.}
  {\bfseries 117} (2016) 182002}
  [\href{https://arxiv.org/abs/1606.02625}{{\ttfamily 1606.02625}}].

\bibitem{Sirunyan:2018nqx}
{\scshape CMS} collaboration, A.~M. Sirunyan et~al., \emph{{Measurement of the
  inelastic proton-proton cross section at $\sqrt{s}=$ 13 TeV}},
  \href{https://arxiv.org/abs/1802.02613}{{\ttfamily 1802.02613}}.

\bibitem{Antchev:2017dia}
{\scshape TOTEM} collaboration, G.~Antchev et~al., \emph{{First measurement of
  elastic, inelastic and total cross-section at $\sqrt{s}=13$ TeV by TOTEM and
  overview of cross-section data at LHC energies}},
  \href{https://arxiv.org/abs/1712.06153}{{\ttfamily 1712.06153}}.

\bibitem{Aad:2012pw}
{\scshape ATLAS} collaboration, G.~Aad et~al., \emph{{Rapidity gap cross
  sections measured with the ATLAS detector in $pp$ collisions at $\sqrt{s}=7$
  TeV}}, \href{https://doi.org/10.1140/epjc/s10052-012-1926-0}{\emph{Eur. Phys.
  J.} {\bfseries C72} (2012) 1926}
  [\href{https://arxiv.org/abs/1201.2808}{{\ttfamily 1201.2808}}].

\bibitem{Ryskin:2011qe}
M.~G. Ryskin, A.~D. Martin and V.~A. Khoze, \emph{{High-energy strong
  interactions: from `hard' to `soft'}},
  \href{https://doi.org/10.1140/epjc/s10052-011-1617-2}{\emph{Eur. Phys. J.}
  {\bfseries C71} (2011) 1617}
  [\href{https://arxiv.org/abs/1102.2844}{{\ttfamily 1102.2844}}].

\bibitem{Cudell:2002xe}
{\scshape COMPETE} collaboration, J.~R. Cudell, V.~V. Ezhela, P.~Gauron,
  K.~Kang, {\relax Yu}.~V. Kuyanov, S.~B. Lugovsky et~al., \emph{{Benchmarks
  for the forward observables at RHIC, the Tevatron Run II and the LHC}},
  \href{https://doi.org/10.1103/PhysRevLett.89.201801}{\emph{Phys. Rev. Lett.}
  {\bfseries 89} (2002) 201801}
  [\href{https://arxiv.org/abs/hep-ph/0206172}{{\ttfamily hep-ph/0206172}}].

\bibitem{Patrignani:2016xqp}
{\scshape Particle Data Group} collaboration, C.~Patrignani et~al.,
  \emph{{Review of Particle Physics}},
  \href{https://doi.org/10.1088/1674-1137/40/10/100001}{\emph{Chin. Phys.}
  {\bfseries C40} (2016) 590}.

\bibitem{Kwiecinski:1980wb}
J.~Kwiecinski and M.~Praszalowicz, \emph{{Three Gluon Integral Equation and Odd
  c Singlet Regge Singularities in QCD}},
  \href{https://doi.org/10.1016/0370-2693(80)90909-0}{\emph{Phys. Lett.}
  {\bfseries 94B} (1980) 413}.

\bibitem{Bartels:1980pe}
J.~Bartels, \emph{{High-Energy Behavior in a Nonabelian Gauge Theory (II)}},
  \href{https://doi.org/10.1016/0550-3213(80)90019-X}{\emph{Nucl. Phys.}
  {\bfseries B175} (1980) 365}.

\bibitem{Bartels:1999yt}
J.~Bartels, L.~N. Lipatov and G.~P. Vacca, \emph{{A New odderon solution in
  perturbative QCD}},
  \href{https://doi.org/10.1016/S0370-2693(00)00221-5}{\emph{Phys. Lett.}
  {\bfseries B477} (2000) 178}
  [\href{https://arxiv.org/abs/hep-ph/9912423}{{\ttfamily hep-ph/9912423}}].

\bibitem{Ryskin:1987ya}
M.~G. Ryskin, \emph{{Odderon and Polarization Phenomena in QCD}}, {\emph{Sov.
  J. Nucl. Phys.} {\bfseries 46} (1987) 337}.

\bibitem{Braun:1998fs}
M.~A. Braun, \emph{{Odderon and QCD}},
  \href{https://arxiv.org/abs/hep-ph/9805394}{{\ttfamily hep-ph/9805394}}.

\bibitem{Ewerz}
C.~Ewerz, \emph{{The Odderon: Theoretical status and experimental tests}},  in
  \emph{{11th International Conference on Elastic and Diffractive Scattering:
  Towards High Energy Frontiers: The 20th Anniversary of the Blois Workshops,
  17th Rencontre de Blois (EDS 05) Chateau de Blois, Blois, France, May 15-20,
  2005}}, 2005, \href{https://arxiv.org/abs/hep-ph/0511196}{{\ttfamily
  hep-ph/0511196}}.

\bibitem{Block}
M.~M. Block, \emph{{Hadronic forward scattering: Predictions for the Large
  Hadron Collider and cosmic rays}},
  \href{https://doi.org/10.1016/j.physrep.2006.06.003}{\emph{Phys. Rept.}
  {\bfseries 436} (2006) 71}
  [\href{https://arxiv.org/abs/hep-ph/0606215}{{\ttfamily hep-ph/0606215}}].

\bibitem{Finkelstein:1989mf}
J.~Finkelstein, H.~M. Fried, K.~Kang and C.~I. Tan, \emph{{Forward Scattering
  at Collider Energies and Eikonal Unitarization of Odderon}},
  \href{https://doi.org/10.1016/0370-2693(89)91697-3}{\emph{Phys. Lett.}
  {\bfseries B232} (1989) 257}.

\bibitem{Khoze:2018bus}
V.~A. Khoze, A.~D. Martin and M.~G. Ryskin, \emph{{Black disk, maximal Odderon
  and unitarity}},
  \href{https://doi.org/10.1016/j.physletb.2018.03.025}{\emph{Phys. Lett.}
  {\bfseries B780} (2018) 352}
  [\href{https://arxiv.org/abs/1801.07065}{{\ttfamily 1801.07065}}].

\bibitem{Antchev:2016vpy}
{\scshape TOTEM} collaboration, G.~Antchev et~al., \emph{{Measurement of
  elastic pp scattering at $\sqrt{\hbox {s}} = \hbox {8}$ TeV in the
  Coulomb–nuclear interference region: determination of the $\mathbf {\rho }$
  -parameter and the total cross-section}},
  \href{https://doi.org/10.1140/epjc/s10052-016-4399-8}{\emph{Eur. Phys. J.}
  {\bfseries C76} (2016) 661}
  [\href{https://arxiv.org/abs/1610.00603}{{\ttfamily 1610.00603}}].

\bibitem{Amos:1991bp}
{\scshape E710} collaboration, N.~A. Amos et~al., \emph{{Measurement of $\rho$,
  the ratio of the real to imaginary part of the $\bar{p} p$ forward elastic
  scattering amplitude, at $\sqrt{s}$ = 1.8-TeV}},
  \href{https://doi.org/10.1103/PhysRevLett.68.2433}{\emph{Phys. Rev. Lett.}
  {\bfseries 68} (1992) 2433}.

\bibitem{Haguenauer:1993kan}
{\scshape UA4/2} collaboration, C.~Augier et~al., \emph{{A Precise measurement
  of the real part of the elastic scattering amplitude at the S anti-p p S}},
  \href{https://doi.org/10.1016/0370-2693(93)90350-Q}{\emph{Phys. Lett.}
  {\bfseries B316} (1993) 448}.

\bibitem{Anselm:1972ir}
A.~A. Anselm and V.~N. Gribov, \emph{{Zero pion mass limit in interactions at
  very high-energies}},
  \href{https://doi.org/10.1016/0370-2693(72)90559-X}{\emph{Phys. Lett.}
  {\bfseries 40B} (1972) 487}.

\bibitem{Khoze:2000wk}
V.~A. Khoze, A.~D. Martin and M.~G. Ryskin, \emph{{Soft diffraction and the
  elastic slope at Tevatron and LHC energies: A MultiPomeron approach}},
  \href{https://doi.org/10.1007/s100520000494}{\emph{Eur. Phys. J.} {\bfseries
  C18} (2000) 167} [\href{https://arxiv.org/abs/hep-ph/0007359}{{\ttfamily
  hep-ph/0007359}}].

\bibitem{Khoze:2014nia}
V.~A. Khoze, A.~D. Martin and M.~G. Ryskin, \emph{{t dependence of the slope of
  the high energy elastic pp cross section}},
  \href{https://doi.org/10.1088/0954-3899/42/2/025003}{\emph{J. Phys.}
  {\bfseries G42} (2015) 025003}
  [\href{https://arxiv.org/abs/1410.0508}{{\ttfamily 1410.0508}}].

\bibitem{Donnachie:1992ny}
A.~Donnachie and P.~V. Landshoff, \emph{{Total cross-sections}},
  \href{https://doi.org/10.1016/0370-2693(92)90832-O}{\emph{Phys. Lett.}
  {\bfseries B296} (1992) 227}
  [\href{https://arxiv.org/abs/hep-ph/9209205}{{\ttfamily hep-ph/9209205}}].

\bibitem{Antchev:2013paa}
{\scshape TOTEM} collaboration, G.~Antchev et~al.,
  \emph{{Luminosity-Independent Measurement of the Proton-Proton Total Cross
  Section at $\sqrt{s}=8$  TeV}},
  \href{https://doi.org/10.1103/PhysRevLett.111.012001}{\emph{Phys. Rev. Lett.}
  {\bfseries 111} (2013) 012001}.

\bibitem{Nemes:2017gut}
F.~J. Nemes, \emph{{Elastic and total cross-section measurements by TOTEM: Past
  and future}}, {\emph{PoS} {\bfseries DIS2017} (2018) 059}.

\bibitem{Aad:2014dca}
{\scshape ATLAS} collaboration, G.~Aad et~al., \emph{{Measurement of the total
  cross section from elastic scattering in pp collisions at $\sqrt{s}=7$ TeV
  with the ATLAS detector}},
  \href{https://doi.org/10.1016/j.nuclphysb.2014.10.019}{\emph{Nucl. Phys.}
  {\bfseries B889} (2014) 486}
  [\href{https://arxiv.org/abs/1408.5778}{{\ttfamily 1408.5778}}].

\bibitem{Aaboud:2016ijx}
{\scshape ATLAS} collaboration, M.~Aaboud et~al., \emph{{Measurement of the
  total cross section from elastic scattering in $pp$ collisions at
  $\sqrt{s}=8$ TeV with the ATLAS detector}},
  \href{https://doi.org/10.1016/j.physletb.2016.08.020}{\emph{Phys. Lett.}
  {\bfseries B761} (2016) 158}
  [\href{https://arxiv.org/abs/1607.06605}{{\ttfamily 1607.06605}}].

\bibitem{Alkin:2014rfa}
A.~Alkin, E.~Martynov, O.~Kovalenko and S.~M. Troshin, \emph{{Impact-parameter
  analysis of TOTEM data at the LHC: Black disk limit exceeded}},
  \href{https://doi.org/10.1103/PhysRevD.89.091501}{\emph{Phys. Rev.}
  {\bfseries D89} (2014) 091501}
  [\href{https://arxiv.org/abs/1403.8036}{{\ttfamily 1403.8036}}].

\end{thebibliography}\endgroup

\enddocument